\documentclass[a4paper,11pt]{article}
\usepackage{pos}
\usepackage{enumitem}
\usepackage{gensymb}
\usepackage{titlesec}
\titlespacing*{\section}
  {0pt}{.25\baselineskip}{.25\baselineskip}
\titlespacing*{\subsection}
  {0pt}{.25\baselineskip}{.25\baselineskip}

\title{The Simons Observatory: Design, Integration, and Current Status of Small Aperture Telescopes}
\ShortTitle{SO SATs: Design, Integration, and Current Status}

\author*[a]{Aashrita Mangu}
\author[a]{Lance Corbett}
\author[b]{Sanah Bhimani}
\author[c]{Fred Carl}
\author[c]{Samuel Day-Weiss}
\author[a]{Brooke DiGia}
\author[d]{Josquin Errard}
\author[e]{Nicholas Galitzki}
\author[f]{Masashi Hazumi}
\author[g, h]{Shawn W. Henderson}
\author[a]{Varun Kabra}
\author[i]{Amber Miller}
\author[j]{Jenna Moore}
\author[a]{Xue Song}
\author[k]{Tran Tsan}
\author[c]{Yuhan Wang}
\author[k]{Andrea Zonca}

\affiliation[a]{University of California Berkeley, Berkeley, CA, USA}

\affiliation[b]{Yale University, New Haven, CT, USA}

\affiliation[c]{Princeton University,Princeton, NJ, USA}

\affiliation[d]{Université Paris Cité, CNRS, Paris, France}

\affiliation[e]{University of Texas at Austin, Austin, TX, USA}

\affiliation[f]{QUP, KEK, Tsukuba, Japan}

\affiliation[g]{Stanford University, Stanford, CA, USA}

\affiliation[h]{SLAC National Accelerator Laboratory, Menlo Park, CA, USA}

\affiliation[i]{University of Southern California, Los Angeles, CA, USA}

\affiliation[j]{Arizona State University, Tempe, AZ, USA}

\affiliation[k]{University of California San Diego, La Jolla, CA, USA}

\emailAdd{amangu@berkeley.edu}

\abstract{The Simons Observatory (SO) is a cosmic microwave background (CMB) survey experiment located in the Atacama Desert in Chile at an elevation of 5200 meters, nominally consisting of an array of three 0.42-meter small aperture telescopes (SATs) and one 6-meter large aperture telescope (LAT). SO will make accurate measurements of the CMB temperature and polarization spanning six frequency bands ranging from 27 to 280 GHz, fielding a total of $\sim$68,000 detectors covering angular scales between one arcminute to tens of degrees. In this paper, we focus on the SATs, which are tailored to search for primordial gravitational waves, with the primary science goal of measuring the primordial tensor-to-scalar ratio \textit{r} at a target level of $\sigma(r) \approx 0.003$. We discuss the design drivers, scientific impact, and current deployment status of the three SATs, which are scheduled to start taking data in the coming year. The SATs aim to map 10\% of the sky at a 2 $\mu$K-arcmin noise level observing at mid-frequencies (93/145 GHz), with additional ultra-high-frequency (225/280 GHz) and low-frequency (27/39 GHz) targets to yield galactic foreground-subtracted measurements. }

\FullConference{XVIII International Conference on Topics in Astroparticle and Underground Physics (TAUP2023)\\
 28.08-01.09.2023\\
University of Vienna\\}

\begin{document}
\maketitle
\section{The Simons Observatory}

Over the past several decades, cosmic microwave background (CMB) experiments have amassed new insights into the fundamental sciences and advanced the standard cosmological model ($\Lambda$CDM) (e.g. \cite{cobe1992}). Experiments like the Simons Observatory (SO) are attempting to find evidence for inflation, a rapid expansion of space-time soon after the universe began, which would help solve issues in the $\Lambda$CDM model and explain the path towards the current structure of the universe \cite{SOscience, cmbs4science}. Polarization signals in the CMB are often written as an effective Helmholtz decomposition into curl-free E and divergence-free B modes. Inflationary B-modes are hypothesized products of primordial gravitational waves. The search for this faint signal requires understanding and removing galactic and extragalactic foregrounds which contribute to the B-mode signal as well. In particular, synchrotron emission at low frequencies and dust emission at high frequencies within our observing windows must be removed. A primordial E-mode signal can also be gravitationally perturbed by large scale structures into a lensing B-mode signal. These lensing B-mode signals contain valuable information about the structure of the universe, but must be subtracted out to extract any inflationary signals. The inflationary B-mode signal can be parameterized by \textit{r}, the tensor-to-scalar ratio of CMB polarization, as an effective energy scale for inflation \cite{Baumann_2009, seljack1997, Kamionkowski_1997, Kamionkowski_2016}.

SO is a CMB experiment currently deploying an array of three 0.42-meter small aperture telescopes (SATs) and one 6-meter large aperture telescope (LAT) in the Atacama Desert in Chile \cite{Xu_2020, Xu_2021, Ali_2020, Galitzki_2018, Kiuchi_2020}. SO will nominally make accurate measurements of the CMB, fielding a total of $\sim$68,000 Transition Edge Sensor (TES) detectors covering angular scales between one arcminute to tens of degrees \cite{Ali_2020, SOscience}. The SATs and the LAT will be observing mid-frequencies (MF) centered at 93 and 145 GHz, ultra-high-frequencies (UHF) centered at 225 and 280 GHz, and low-frequencies (LF) centered at 27 and 39 GHz. We are currently deploying the LAT, two MF SATs, and one UHF SAT. SO will search for inflationary signatures and study particle physics beyond the Standard Model by constraining the sum of neutrino masses, probing the nature of dark energy and dark matter, and enabling a model-independent search for light relic particles from the early universe. The SATs are tailored to search for primordial gravitational waves, with the primary science goal of measuring \textit{r} at a target level of $\sigma(r) \approx 0.003$ \cite{planck2018_1, SOscience}. In this paper, we discuss the SO SAT design, integration and testing, and current status of deployment.

\section{SAT Design}

\begin{figure}[t]
\centerline{\includegraphics[width=\textwidth]{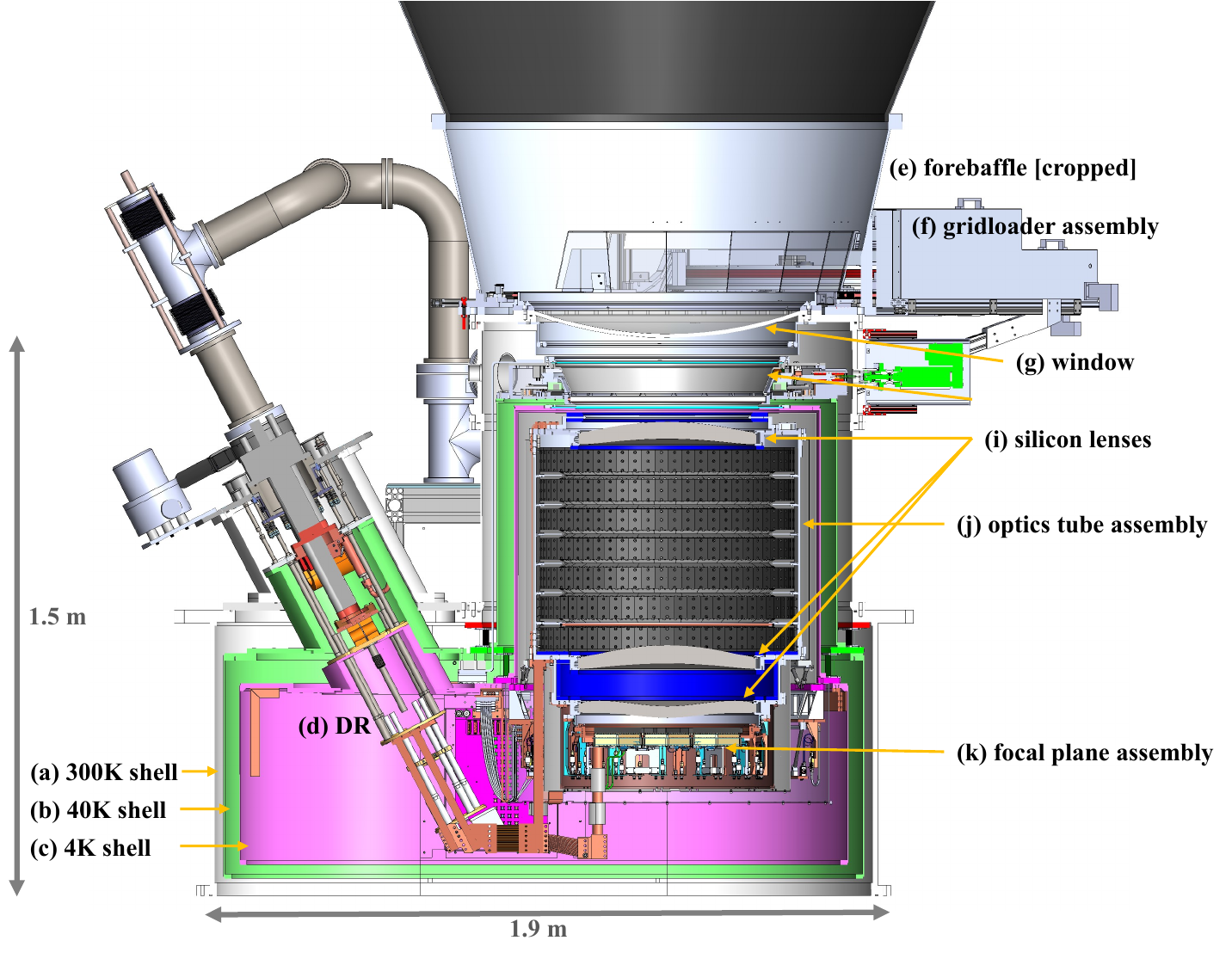}}
\caption{\small{SAT structure, cross sectional view. Forebaffle structure is cropped in the above image. Not labelled explicitly are the series of low pass and alumina filters located at 40K, 4K, 1K, and 100mK.}}
\label{fig:sat}
\end{figure}

%\begin{figure}[t]
%\centerline{\includegraphics[width=\textwidth]{figures/sat_backend_labelled.pdf}}
%\caption{\small{SAT structure, backend view.}}
%\label{fig:satback}
%\end{figure}

%\begin{figure}[t]
%\centerline{\includegraphics[width=\textwidth]{figures/sat_combined.pdf}}
%\caption{\small{SAT structure, cross sectional view. Forebaffle structure is cropped in the above image. Not labelled explicitly are the series of low pass and alumina filters located at 40K, 4K, 1K, and 100mK.}}
%\label{fig:sat}
%\end{figure}

\begin{figure}[t]
\centerline{\includegraphics[width=\textwidth]{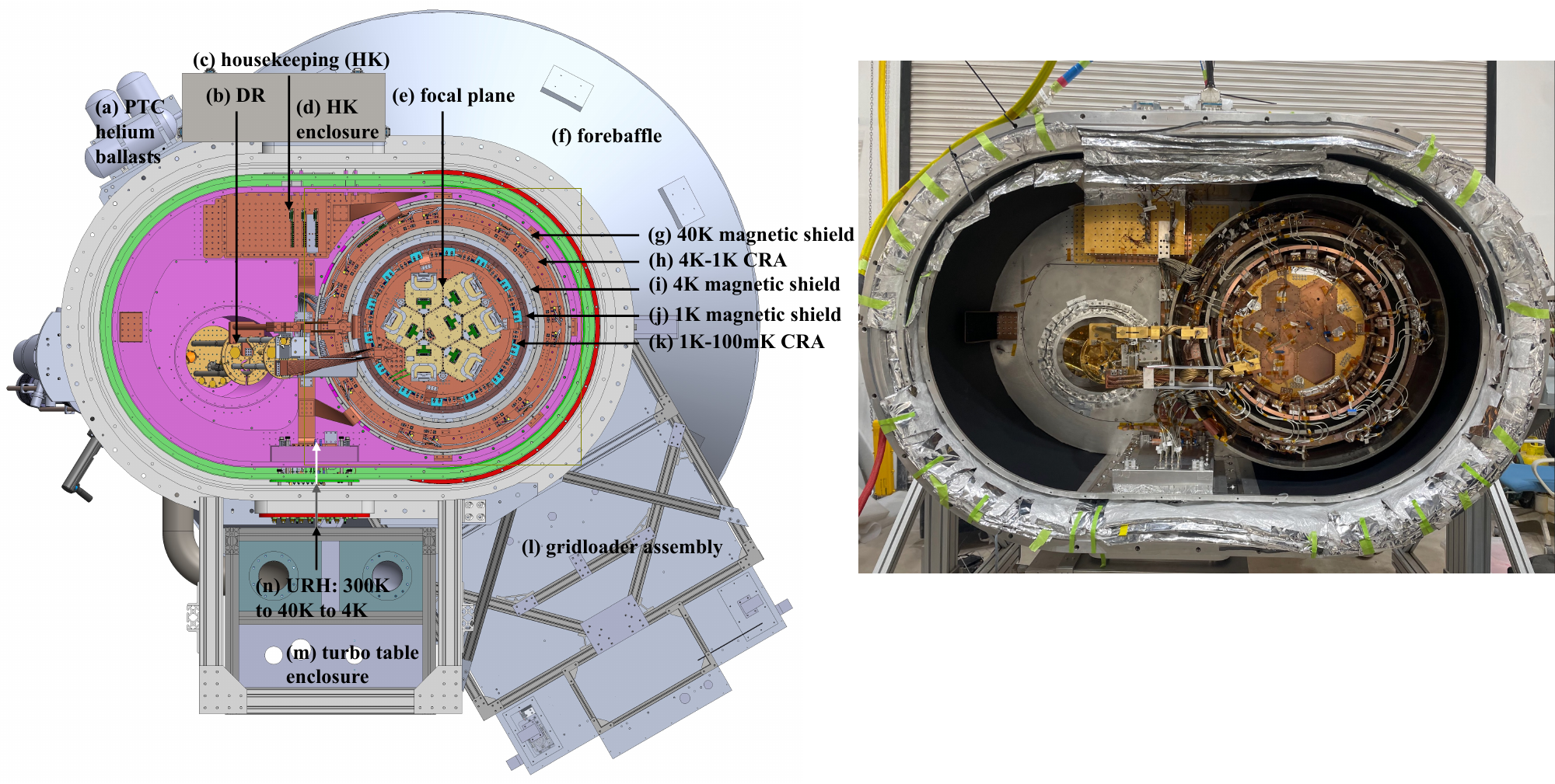}}
\caption{\small{SAT structure, backend view. Left is CAD of full assembly. Right is SAT-MF1 integration in the high bay at the Chilean site, prior to being placed on the SAT platform.}}
\label{fig:satback}
\end{figure}

All SATs contain the same basic structure, as shown in Figures \ref{fig:sat} and \ref{fig:satback}. The major cryogenic components include a Pulse Tube Cooler (PTC) and Dilution Refrigerator (DR). The DR system comes with its own additional PTC. The two PTCs help get the cryostat to 4K, and the DR allows the SAT optics tube to sit at 1K and focal plane to sit at $\sim$100mK. 

The SAT itself has a $\sim$35$\degree$ field-of-view. The forebaffle encompassing a volume above the window reflects radiation incident at angles greater than $\sim$40$\degree$ off-axis out of the telescope to reduce loading into the system. The gridloader assembly loads a sparse wire grid to allow precise polarization calibration with a detector angle systematic error of 0.08$\degree$ \cite{gridloader2023}. The optical path of a CMB photon can be followed from the window all the way down to the focal plane. 

A photon interacts with a cryogenic continuously rotating cold half-wave plate (CHWP), a series of low pass and alumina filters, and 3 lenses to be focused onto the focal plane \cite{Kiuchi_2020, Galitzki_2018, Ali_2020}. The CHWP is a magnetically levitating, rotating 3-layer birefringent (sapphire) polarization modulator. It reduces low frequency atmospheric noise and mitigates systematic effects that would otherwise occur from pair-differencing orthogonal detectors \cite{yamada2023, Hill_2020}. The focal plane in the SATs and LAT are hexagonally tiled with Universal Focal Plane Modules (UFMs), which house the dual-polarized, dichroic TESs \cite{mccarrick_2021mf, Healy_2022uhf}. These UFMs consist of both optical components and readout components \cite{McCarrick_2021, Healy_2022}. SO uses microwave multiplexing, which allows \textit{O}(1000) detectors per readout chain. Each readout chain traverses a cold readout assembly (CRA) and universal readout harness (URH) \cite{Rao_2020, moore2022development}. The CRA takes the signal from 100mK to 4K, and the URH takes the signal from 4K to the 300K readout port on the SAT. This signal is amplified at the 4K and 40K stages and demodulated with the warm readout electronics developed by SLAC (i.e. SLAC Microresonator Radio Frequency or SMuRF electronics) \cite{Henderson_2018, Yu_2023}. The focal plane is protected from magnetic fields by a series of magnetic shields at the 1K, 4K, and 40K stages \cite{Galitzki_2018, Ali_2020, Kiuchi_2020}. Finally, the thermal environment of the SAT is tracked through the housekeeping port, which reads out all diodes and ruthenium oxide sensors strategically placed around the SAT. 

In an effort to streamline design and production of the SO telescopes, there are some key components that are designed to be interchangeable between the LAT and any of the SATs. For example, any component labelled with `Universal' has this characteristic like the URH and UFMs (which also consist of a Universal Multiplexing Module attached to the optical stack) \cite{Healy_2022, Healy_2022uhf, mccarrick_2021mf, McCarrick_2021}. Similarly, the SMuRF system is used across all telescopes. Conversely, there are some non-negligible differences between the SATs. Different frequency bands require different optics: specifically, the filters, filter thicknesses, lenses, CHWP sapphire stack thickness, and different detector parameters (which live in the UFMs). Also, while SAT-MF1 and SAT-UHF use mullite-duroid anti-reflection coatings, SAT-MF2 uses metamaterial coatings. Small mechanical differences between all SATs are expected, and are taken into account for integration and testing requirements.

\section{Integration and Testing Strategy}

%\begin{figure}[t]
%\centerline{\includegraphics[width=\textwidth]{figures/sat_assembled.pdf}}
%\caption{\small{SAT assembly. \textbf{(a)} SAT-UHF focal plane for final in-lab testing. \textbf{(b)} Universal Focal Plane Modules (UFMs) at Chilean site before being placed in SAT-MF1. \textbf{(c)} SAT-MF1 integration in the highbay at Chilean site prior to being closed up and placed on the SAT platform.}}
%\label{fig:satassembled}
%\end{figure}

Every SAT has been carefully integrated and tested in steps to ensure deployment readiness. SO is an important precursor to experiments like CMB-S4 to ensure technological readiness and resource allocation. Such large scale experiments involve deployment strategies to get many telescopes to the field. For SO, all 3 SATs have the same basic structure. Thus, it was prudent to stagger integration of the SATs sequentially to match hardware production timelines. Any design changes that were needed from testing the first SAT (SAT-MF1) trickled down to SAT-MF2 and SAT-UHF to save integration and testing time on the following SATs. Here we provide a brief outline of testing for deployment readiness:
\begin{itemize}[nosep,leftmargin=*]
    \item \textbf{Mechanical:} Thermal and vacuum performance, CHWP mechanical performance, grid loader and rotator mechanical performance, magnetic performance, equipment enclosures
    \item \textbf{Electrical:} Readout performance, detector performance, cryogenic thermal sensing and control, warm housekeeping sensor performance
    \item \textbf{Optical:} Optical component performance, spectral performance, near and far-field optical response, end-to-end optical efficiency, polarization performance
    \item \textbf{Safety:} Motion controllers and actuators, integrated system safety
    \item \textbf{Site Preparation:} Commissioning and operations software, packing and shipping plan, deployment plan
\end{itemize}

SAT-MF1 did a thorough review of all deployment requirements and retired many risks for the other SATs. As a result, SAT-MF2 and SAT-UHF were able to reach deployment readiness faster. For example, grid loader mechanical performance became a retired risk. This strategy also allowed SAT-MF2 and SAT-UHF to diverge from the SAT-MF1 testing plans and tackle new problems. For example, SAT-MF2 was able to test the CHWP sag at different elevation angles, and SAT-UHF was able to test truss strength and search for high frequency leaks in its optical chain. Thus, the later SATs could perform focused studies to address new concerns and formulate a more comprehensive picture of the telescope operation. On the software side, much of the pipeline and analysis development was pushed to near-completion for SAT-MF1. Thus, deployment readiness review of certain software such as commissioning and operations software were not repeated for other SATs, and were instead further developed directly at the Chilean site. 

While SAT-MF1 was reviewed on all the above topics, the last deploying nominal SAT (SAT-UHF) was reviewed on the following condensed version of this: thermal and vacuum performance, CHWP mechanical performance, equipment enclosure status, readout performance, detector performance, optical component performance, personnel and instrument safety, and a combined shipping and deployment plan. Basic mechanical, electrical, and optical performance across all SATs was ensured prior to packing approval. For site, each SAT had to develop an independent packing and shipping plan along with a deployment plan. Staggered deployment at the Chilean site allows solutions of new issues to trickle down to the other SATs similar to the in-lab integration and testing strategy.

\section{Current Status and Future Developments}

SAT-MF1 is currently on the platform at the Chilean site and is on track for starting science observations in 2024. Reassembly, integration, and testing in a highbay at site was done prior to placing the SAT on the platform. SAT systematics testing is currently underway prior to full science observation. As described earlier, SAT-MF1 retired many site-related risks and enabled the quick integration of SAT-MF2. Now that much of the site infrastructure is in place from SAT-MF1 integration, SAT-MF2 is already on the platform as well. SAT-UHF is currently at Lawrence Berkeley National Laboratory and will be packing and shipping for Chile by the end of November 2023. The final in-lab cooldown was a success, and the team is working towards site preparation. All SATs are thus on track for the 2024 science observations goal.

%Full SO will consist of the currently deploying nominal SO, Advanced SO (ASO) [CITE], SO:UK, and SO:Japan [CITE BOTH]. ASO will include 6 new optics tubes (OT) in the LAT such that all 13 OTs are fully populated. Without any further telescope development, this will thus effectively doubling the mapping speed of the LAT for delensing and other science, and also enable transient science. ASO will also include significant data management advances, and replace 70\% of power at site with solar energy using photovoltaic arrays. ASO will begin science observations in 2028. SO:UK and SO:Japan are also fully funded to double the number of SATs on the sky in Chile by 2026. SO:UK will deploy one MF and one UHF SAT using Microwave Kinetic Inductance Detectors (MKIDs) instead of traditionally used TES detectors. SO:Japan will deploy the first LF SAT.

\section{Conclusion}

We reviewed the design, integration, testing and current status of the SO SATs, which are currently in a deployment phase. Nominal SO deployment will begin full observations in 2024. The SATs will enable the high sensitivities needed to search for inflationary signals with the target level of $\sigma(r) \approx 0.003$. There are also plans to reach for higher sensitivities by adding an additional LF, MF, and UHF SAT in the next few years. The LF and UHF SATs will allow foreground cleaning from the primary CMB MF SATs, aided by the LAT which will also support both delensing and foreground subtraction efforts. Updated forecasting and instrument papers are in progress from the collaboration.

\acknowledgments This work was supported in part by the Simons Foundation (Award \#457687, B.K.).

%\begin{thebibliography}{99}
%\bibitem{...}
\bibliographystyle{JHEP}
\bibliography{bib}

%\end{thebibliography}

\end{document}